\documentclass{elsart}
\usepackage{amssymb}
\usepackage{amsfonts}
\usepackage{graphicx}
\usepackage[square,comma]{natbib}

\journal{APFA6 Proceedings}

\begin{document}

\begin{frontmatter}
\title{On the Topological Properties of the World Trade Web: A Weighted Network Analysis.}

\author[Fagiolo]{Giorgio Fagiolo\corauthref{cor}},
\corauth[cor]{Corresponding author. Mail address: Sant'Anna School
of Advanced Studies, Piazza Martiri della Libert\`{a} 33, I-56127
Pisa, Italy. Tel: +39-050-883343; Fax: +39-050-883344. E-mail:
giorgio.fagiolo@sssup.it} \ead{giorgio.fagiolo@sssup.it}
\author[Reyes]{Javier Reyes},
\ead{JReyes@walton.uark.edu}
\author[Schiavo]{Stefano Schiavo}
\ead{stefano.schiavo@ofce.sciences-po.fr}

\address[Fagiolo]{Laboratory of Economics and Management,\\Sant'Anna
School of Advanced Studies, Pisa, Italy. }

\address[Reyes]{Department of Economics, Sam M. Walton College of
Business, \\University of Arkansas, USA.}

\address[Schiavo]{Observatoire Fran\c{c}ais des Conjonctures
\'{E}conomiques,\\
D\'{e}partement de Recherche sur l'Innovation et la Concurrence,
Valbonne, France.}

\begin{abstract}
This paper studies the topological properties of the World Trade Web
(WTW) and its evolution over time by employing a weighted network
analysis. We show that the WTW, viewed as a weighted network,
displays statistical features that are very different from those
obtained by using a traditional binary-network approach. In
particular, we find that: (i) the majority of existing links are
associated to weak trade relationships; (ii) the weighted WTW is
only weakly disassortative; (iii) countries holding more intense
trade relationships are more clustered.
\end{abstract}

\end{frontmatter}

\section{Introduction}
In the last years, a large number of contributions have empirically
explored the topological properties of many real-world networks
\cite{AlbertBarabasi2002,DoroMendes2003,Newman2003,PastosVespignani2004}.
Within this exploding body of literature, econophysicists have
devoted a considerable attention to the World Trade Web (WTW), see
Refs. \cite{SeBo03,Garla2004,Garla2005,LiC03}. In these studies, the
WTW is defined as the network of world-trade relations where
countries play the role of nodes and a link between any two
countries is in place if and only if there exists a non-zero (or a
sufficiently intense) import/export flow between them in a given
year. The picture stemming from these empirical investigations can
be summarized as follows \cite{SeBo03,Garla2004,Garla2005}. First,
the WTW is characterized by a disassortative pattern: countries with
many trade partners are on average connected with countries holding
few partners. Second, partners of well-connected countries are less
interconnected (among themselves) than those of poorly-connected
ones, implying some hierarchical arrangements. Third, the structural
properties of the WTW have remained remarkably stable over time. In
terms of network statistics, two main stylized facts seems therefore
to robustly emerge across the years: (SF1) node degree \citep[see
Ref.][p. 49]{AlbertBarabasi2002} and average nearest-neighbor degree
\citep{Pastor2001} are negatively correlated; (SF2) node degree and
clustering coefficient \citep{WattsStrogatz1998} are negatively
correlated.

To a large extent, however, SF1-2 refer to a binary-network analysis
(BNA).\footnote{See \citep{LiC03} for an exception.} Indeed, Refs.
\cite{SeBo03,Garla2004,Garla2005} only study the WTW as a network
where each link from country $i$ to country $j$ either exists or
not. A binary network can thus be characterized by a binary
adjacency matrix $A$, whose generic entry $a_{ij}=1$ if and only if
a link from node $i$ to $j$ is in place.\footnote{Self-loops, i.e.
links connecting $i$ with itself are not typically considered. This
means that $a_{ii}=0$, for all $i$.} A BNA treats all links in the
WTW as they were completely homogeneous. This is counterintuitive,
as actual import-export flows greatly differ both when their are
evaluated in their levels (e.g. in current U.S. dollars) and when
they are computed as shares of the importing/exporting country size
(measured e.g. by its gross domestic product, GDP). In order to take
into account the existing heterogeneity in the capacity and
intensity of connections, a weighted-network analysis (WNA) can
instead be performed \cite{Barr04,Barr05,Bart05,DeMontis2005}. More
formally, in a WNA each existing link is assigned a value $w_{ij}>0$
proportional to the weight of that link. Hence, a \textit{weighted}
network is fully described by its $N\times N$ weight matrix
$W=\{w_{ij}\}$, where $w_{ii}=0$, all $i$.

In this paper, we explore the statistical properties of the WTW
using a WNA. We ask whether the two stylized facts above still hold
when one weights each existing link with some proxy of the actual
trade flow flowing through it. Our results show that SF1-2 are not
robust to a WNA. In particular, the WTW viewed as a weighted network
is only weakly disassortative. Furthermore, better connected
countries tend to be \textit{more} clustered. The only statistical
feature which resists in a WNA is the constancy over time of WTW
properties. This casts some doubts on whether any process of
globalization (however defined) had some impact on international
trade. The rest of the paper is organized as follows. In Section
\ref{Section:Data} we describe the data and we define network
statistics. Section \ref{Section:Results} discusses our main
results. Finally, Section \ref{Section:Conclusions} concludes.

\section{Data and Network Statistics} \label{Section:Data}
We employ international trade data provided by \cite{GledData2002}
to build a time-sequence of weighted directed networks \citep[see
Ref.][for details]{FRS2007wp}. Our sample refers to $T=20$ years
(1981-2000) and $N=159$ countries. For each country and year, data
report trade flows in current US dollars. To build adjacency and
weight matrices, we followed the flow of goods. This means that rows
represent exporting countries, whereas columns stand for importing
countries. We define a ``trade relationship'' by setting the generic
entry of the adjacency (binary) matrix $\tilde{a}_{ij}^t=1$ if and
only if exports from country $i$ to country $j$ (labeled by
$e_{ij}^t$) are strictly positive in year $t$. Link weights are
instead defined as $\tilde{w}_{ij}^t=e_{ij}^t/GDP_i^t$, i.e. exports
over GDP of the exporting country.\footnote{All our results are
robust to alternative weighting schemes. For example, we
experimented with weights defined as $e_{ij}^t/GDP_j^t$, or simply
as $e_{ij}^t$.} For any particular choice of the weighting setup, we
end up with a sequence of $N\times N$ adjacency and weight matrices
$\{\tilde{A}^t,\tilde{W}^t\}$, $t=1981,...,2000$, which fully
describe the evolution of the WTW from a binary and weighted
directed perspective.

A preliminary statistical analysis of both binary and weighted
matrices suggests that $(\tilde{A}^t,\tilde{W}^t)$ are sufficiently
symmetric to justify an undirected analysis \citep[see Refs.][for
details]{FRS2007wp,FagioloSymmEcoBull}. Therefore, we define entries
of the symmetrized adjacency matrix $a_{ij}=1$ if and only if either
$\tilde{a}_{ij}=1$ or $\tilde{a}_{ji}=1$ (and zero otherwise).
Accordingly, the generic entry of the symmetrized weight matrix $W$
is defined as
$w_{ij}^t=\frac{1}{2}(\tilde{w}_{ij}^t+\tilde{w}_{ji}^t)$. Finally,
in order to have $w_{ij}^t\in[0,1]$ for all $(i,j)$ and $t$, we
renormalize all entries in $W$ by their maximum value.

In this paper, we present results concerning three network
statistics \cite{AlbertBarabasi2002,Pastor2001,WattsStrogatz1998,
DeMontis2005,Saramaki2006,FagioloClustArxiv}. First, for any node
$i$, we compute its \textit{node degree}, defined as
$ND_i=A_{(i)}\textbf{1}$, where $X_{(i)}$ is the $i$-th row of $X$
and $\textbf{1}$ is a unary vector. Node degree can be naturally
extended to weighted networks by computing node strength
$NS_i=W_{(i)}\textbf{1}$. While ND tells us how many partners a node
holds, NS gives us an idea of how intense these relationships are.
Second, we define \textit{average nearest-neighbor degree} of a node
as $ANND_i=(A_{(i)}A\textbf{1})/(A_{(i)}\textbf{1})$. In the case of
weighted networks, this indicator becomes the average
nearest-neighbors strength and is defined as
$ANNS_i=(A_{(i)}W\textbf{1})/(A_{(i)}\textbf{1})$. ANND measures the
average number of partners of a given node's partners, while ANNS
tells us how intense are the relationships maintained by the
partners of a given node. Third, we employ the \textit{clustering
coefficient}, defined for binary networks as
$BCC_i=(A^3)_{ii}/(ND_i(ND_i-1))$ and for weighted networks as
$WCC_i=(W^{\left[\frac{1}{3}\right]})_{ii}^{3}/(ND_i(ND_i-1))$. Here
$(A^3)_{ii}$ is the $i$-th entry on the main diagonal of $A\cdot
A\cdot A$ and $W^{\left[\frac{1}{3}\right]}$ stands for the matrix
obtained from $W$ after raising each entry to $1/3$ \citep[see
Ref.][for a discussion]{FagioloClustArxiv}. BCC counts the fraction
of a node's partners that are themselves partners, while WCC
measures how much intense are the interactions among three intensely
connected partners.

\section{Results} \label{Section:Results}
\begin{figure}[ht]
    \centering
    \begin{minipage}[t]{6cm}
    \includegraphics[width=6cm]{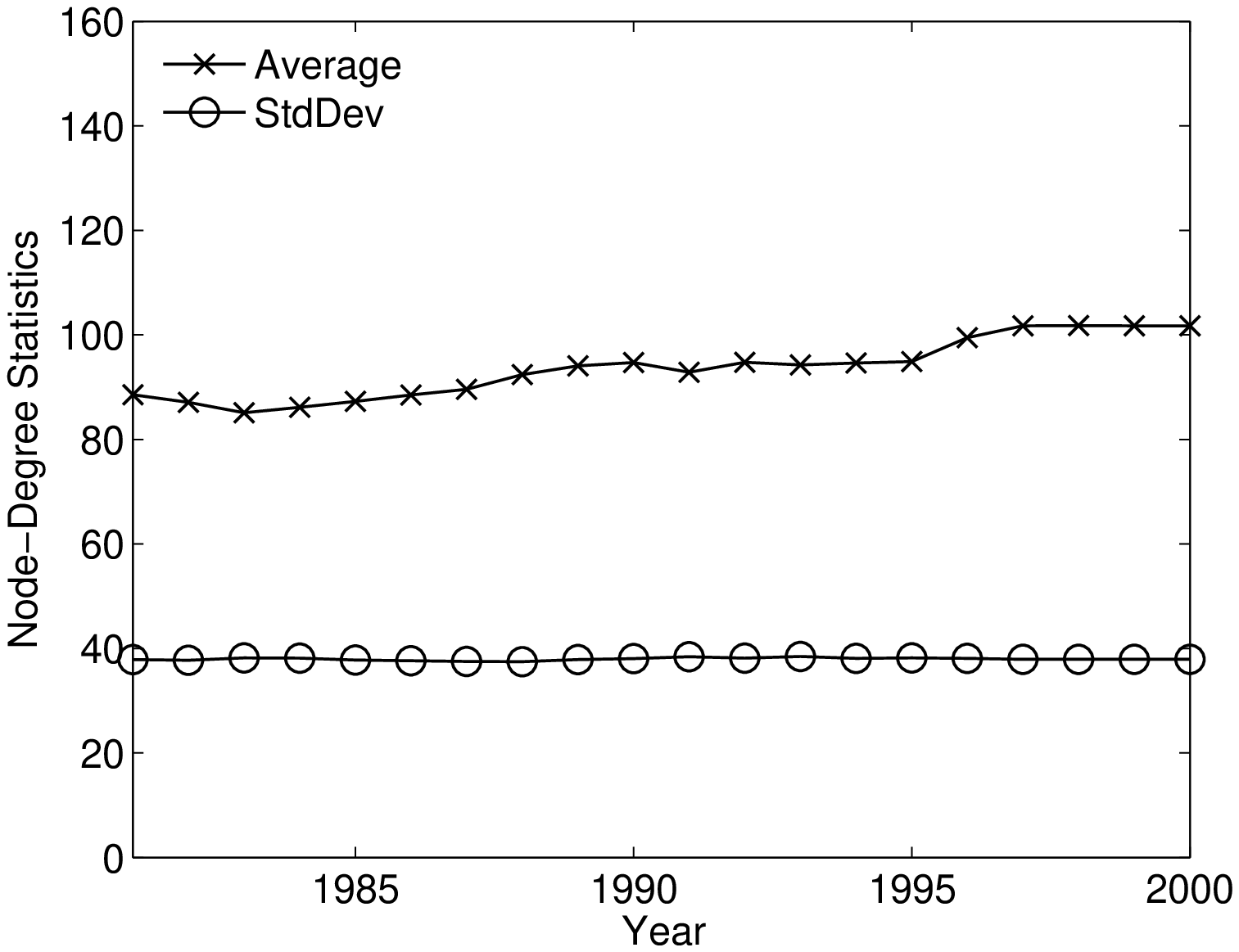}
    \end{minipage}
    \begin{minipage}[t]{6cm}
    \includegraphics[width=6cm]{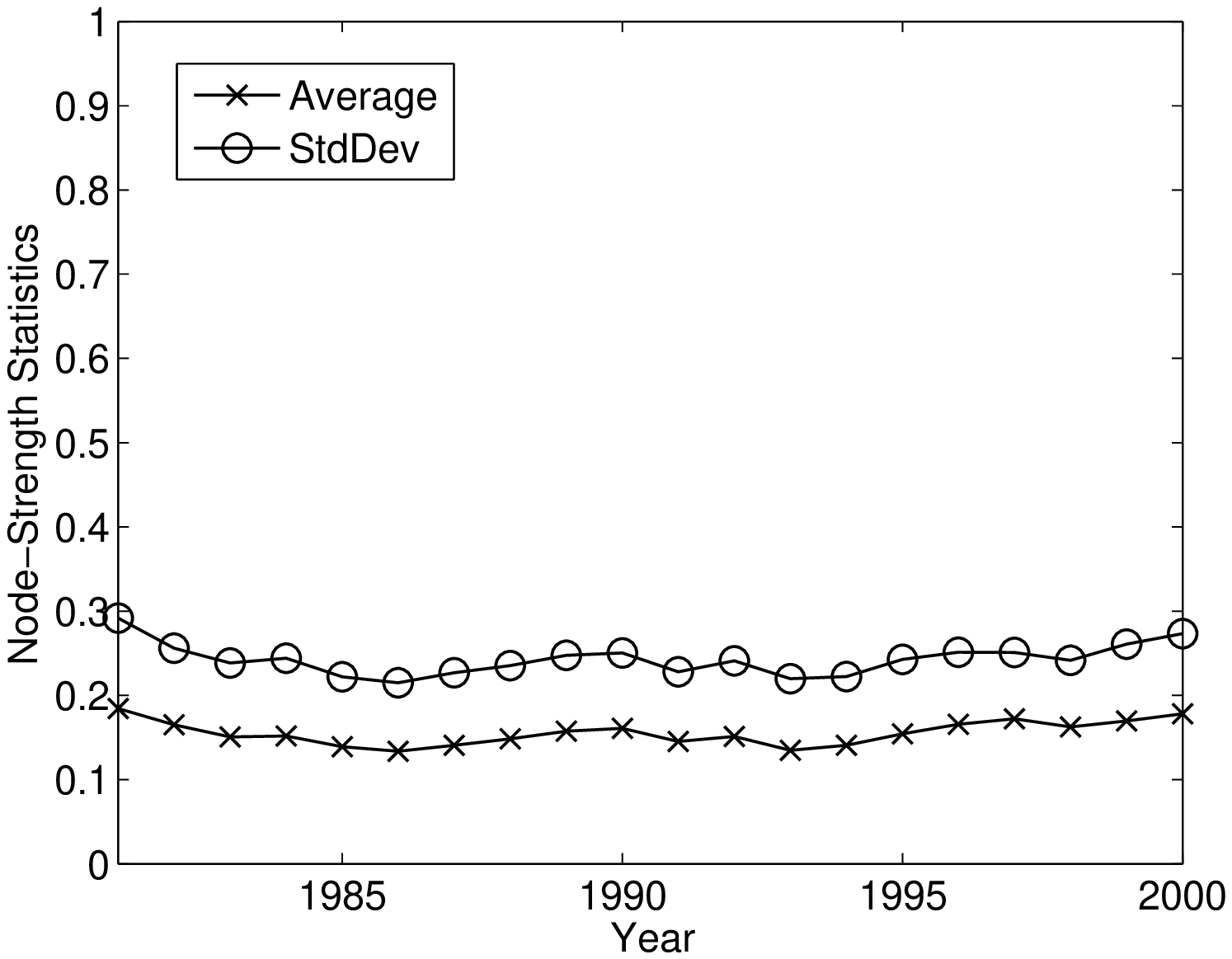}
    \end{minipage}
    \caption{Average and standard deviation of average node degree (left)
    and node strength (right) vs. years.}
    \label{Fig:Ave_Std_ND_NS}
    \end{figure}

We begin by investigating the behavior of ND and NS distributions.
As Figure \ref{Fig:Ave_Std_ND_NS} shows, the WTW as a binary network
is very densely connected. On average each country holds about 90
trade partners (over a maximum of 159). Conversely, the weighted WTW
displays, on a [0,1] scale, a relatively low average NS. Notice also
that the first two moments of both ND and NS have remained
relatively stable over time (if any, average ND has been slowly
increasing). This is a general finding: the first four moments of
all three indicators discussed in Section \ref{Section:Data}, and
their correlations, display a marked time-stationarity. This implies
that the structural properties of the WTW, viewed either as a binary
or as a weighted network, have not been influenced by the process of
globalization (however this may be defined), and confirms the
results in Ref. \cite{Garla2005}. Given this time stationarity, in
the rest of the paper we will focus on a representative year (2000).

\begin{figure}[ht]
    \centering
    \begin{minipage}[t]{6cm}
    \includegraphics[width=6cm]{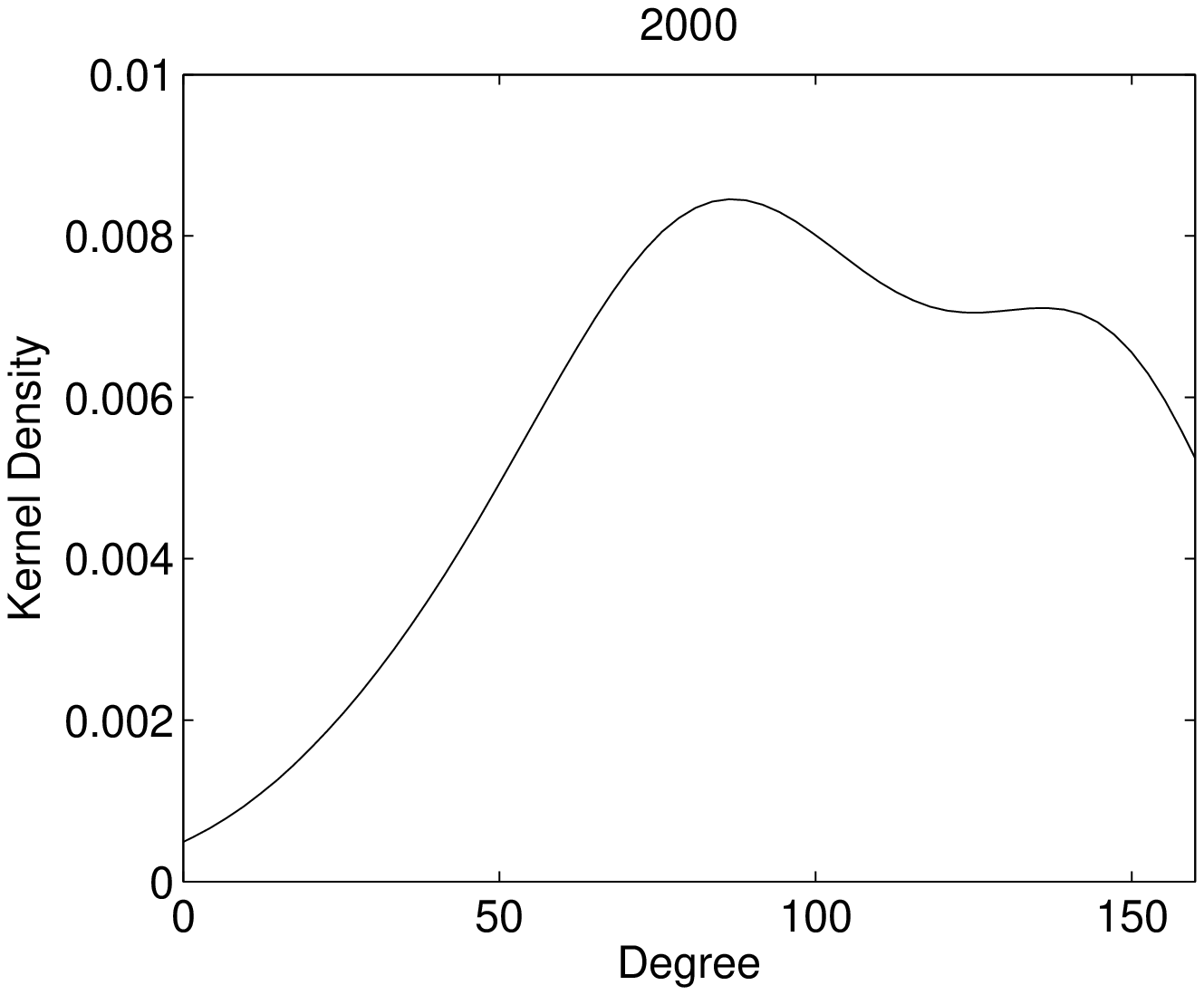}
    \end{minipage}
    \begin{minipage}[t]{6cm}
    \includegraphics[width=6cm]{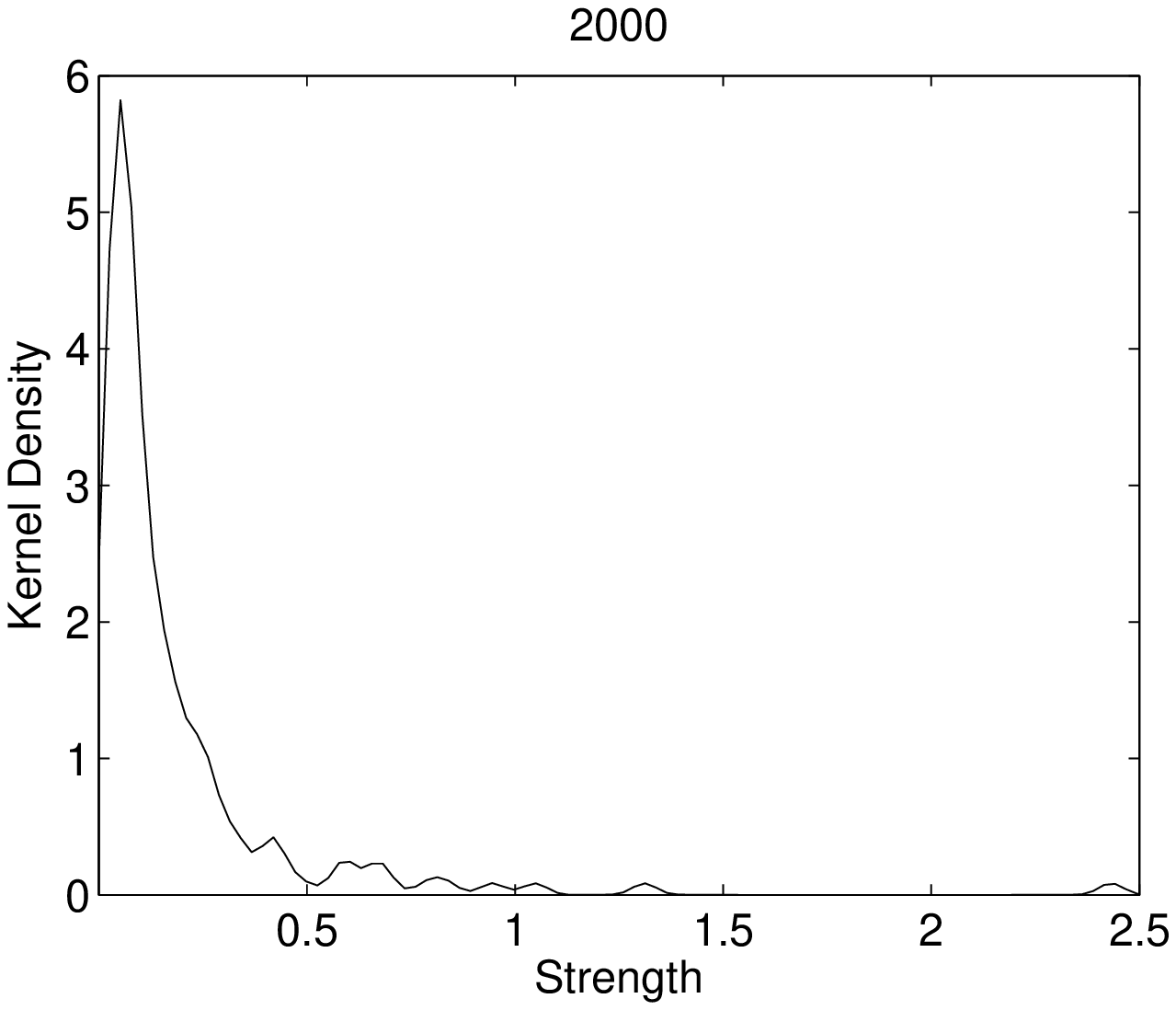}
    \end{minipage}
    \caption{Kernel-smoothed density of node degree (left) and node strength (right)
    distributions. Year: 2000.}
    \label{Fig:Density_ND_NS}
    \end{figure}

The fact that average ND is relatively high, whilst average NS is
low, suggests that the majority of existing connections are
relatively weak. Indeed, the ND-NS correlation coefficient for the
period 1981-2000 is on average 0.50. The structural difference
between ND and NS can be better grasped by plotting the
kernel-smoothed ND and NS density, see Figure
\ref{Fig:Density_ND_NS} for year 2000. The ND distribution is
relatively left-skewed, with a modal value around 90. However, there
exists a group of countries that trade with almost everyone else in
the sample (hence, the second peak around 150).

This picture changes substantially in the weighted case. The NS
distribution is in fact left-skewed: many weak trade relationships
coexist with a few strong ones. Size-rank plots show that NS
distributions are in fact log-normal in the body and Pareto in the
upper tail. This result indicates the presence of a relevant
heterogeneity in the intensity of trade interactions and suggests
that a WNA can provide a more complete description of the
topological properties of the WTW with respect to a BNA.

\begin{figure}[ht]
    \centering
    \begin{minipage}[t]{6cm}
    \includegraphics[width=6cm]{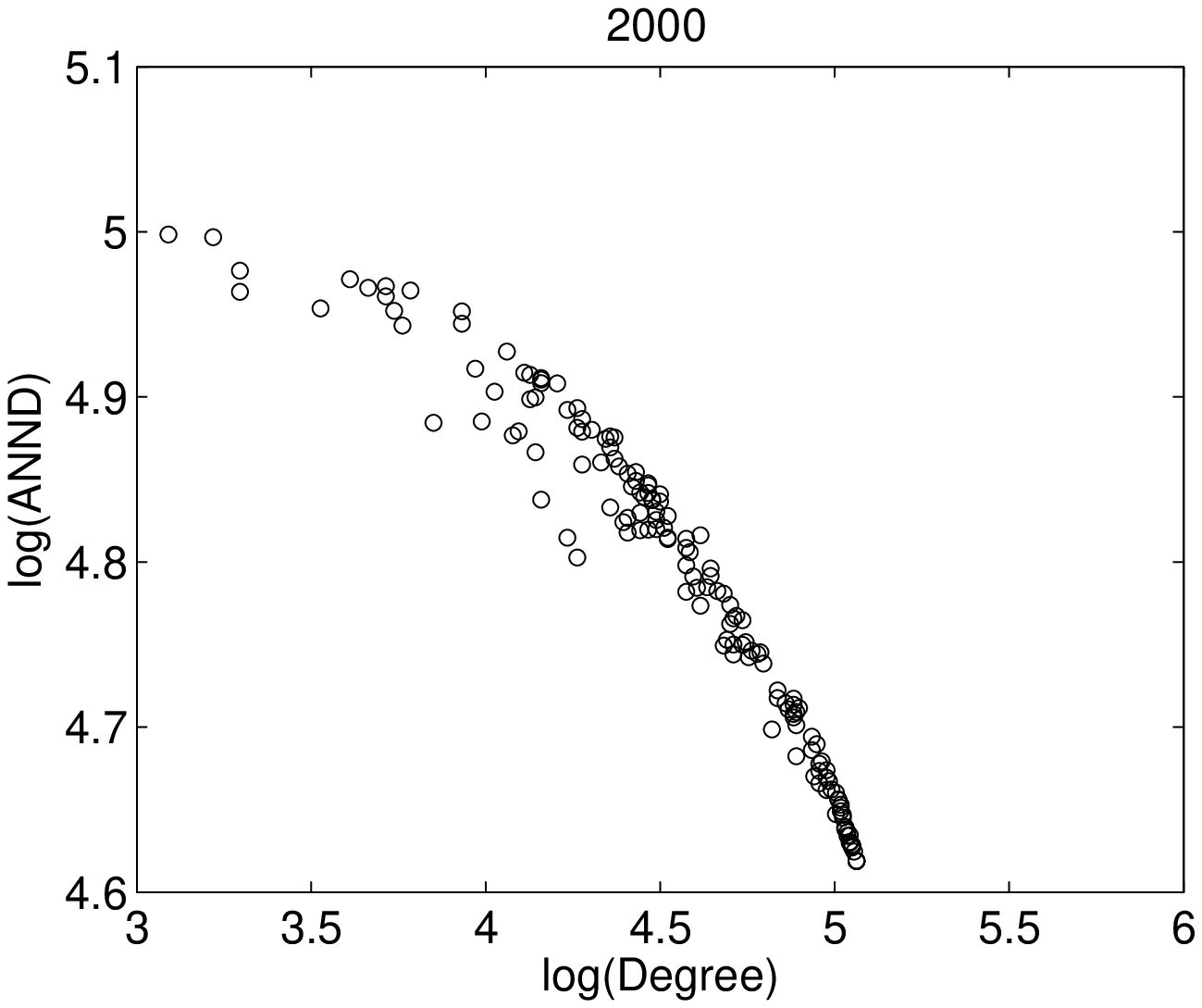}
    \end{minipage}
    \begin{minipage}[t]{6cm}
    \includegraphics[width=6cm]{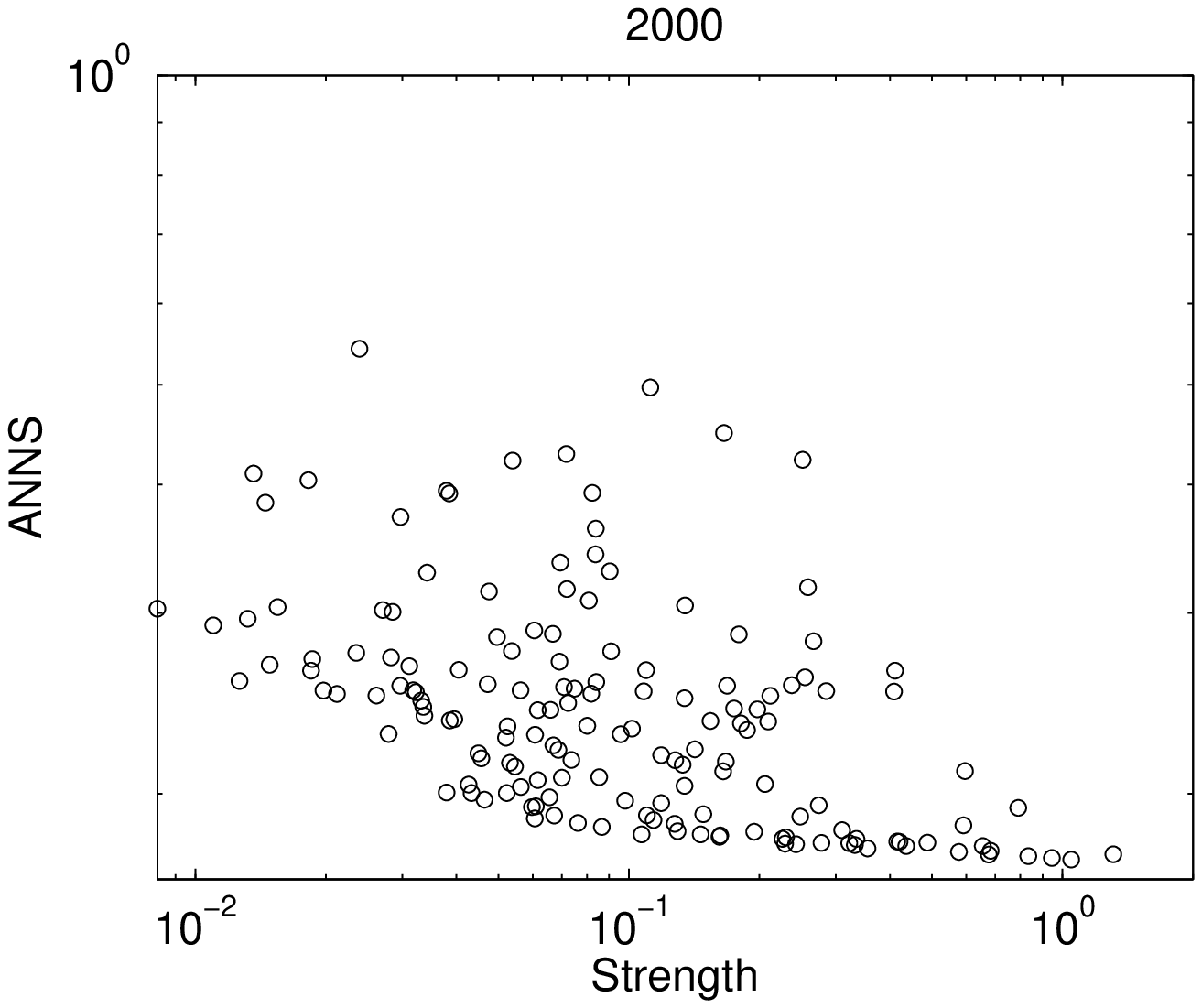}
    \end{minipage}
    \caption{Assortativity analysis. Left: ANND vs. node degree (in logs).
    Right: ANNS vs. node strength (axes in log scale). Year: 2000.}
    \label{Fig:Assortativity}
    \end{figure}

Let us now turn to SF1. As shown in Refs. \cite{SeBo03,Garla2005},
the WTW seems to display a disassortative pattern: countries with
many trade partners are on average connected with countries holding
few partners. This is confirmed in our data. The ND-ANND correlation
coefficient is on average -0.95 across the years. A scatter plot for
year 2000 shows how strong the disassortative pattern is, see Figure
\ref{Fig:Assortativity}, left panel. However, when we plot for the
same year ANNS vs. NS, the correlation pattern appears to be
substantially weaker, cf. Figure \ref{Fig:Assortativity}, right
panel. Countries with medium-low NS are in fact characterized by a
wide range of ANNS values, meaning that there can be well-connected
countries that trade with partners that are also well-connected.
Indeed, the ANNS-NS correlation coefficient stays around -0.40 for
the whole period. Therefore, SF1 does not seem to hold that robustly
when the WTW is studied from a weighted-network perspective.

\begin{figure}[ht]
    \centering
    \begin{minipage}[t]{6cm}
    \includegraphics[width=6cm]{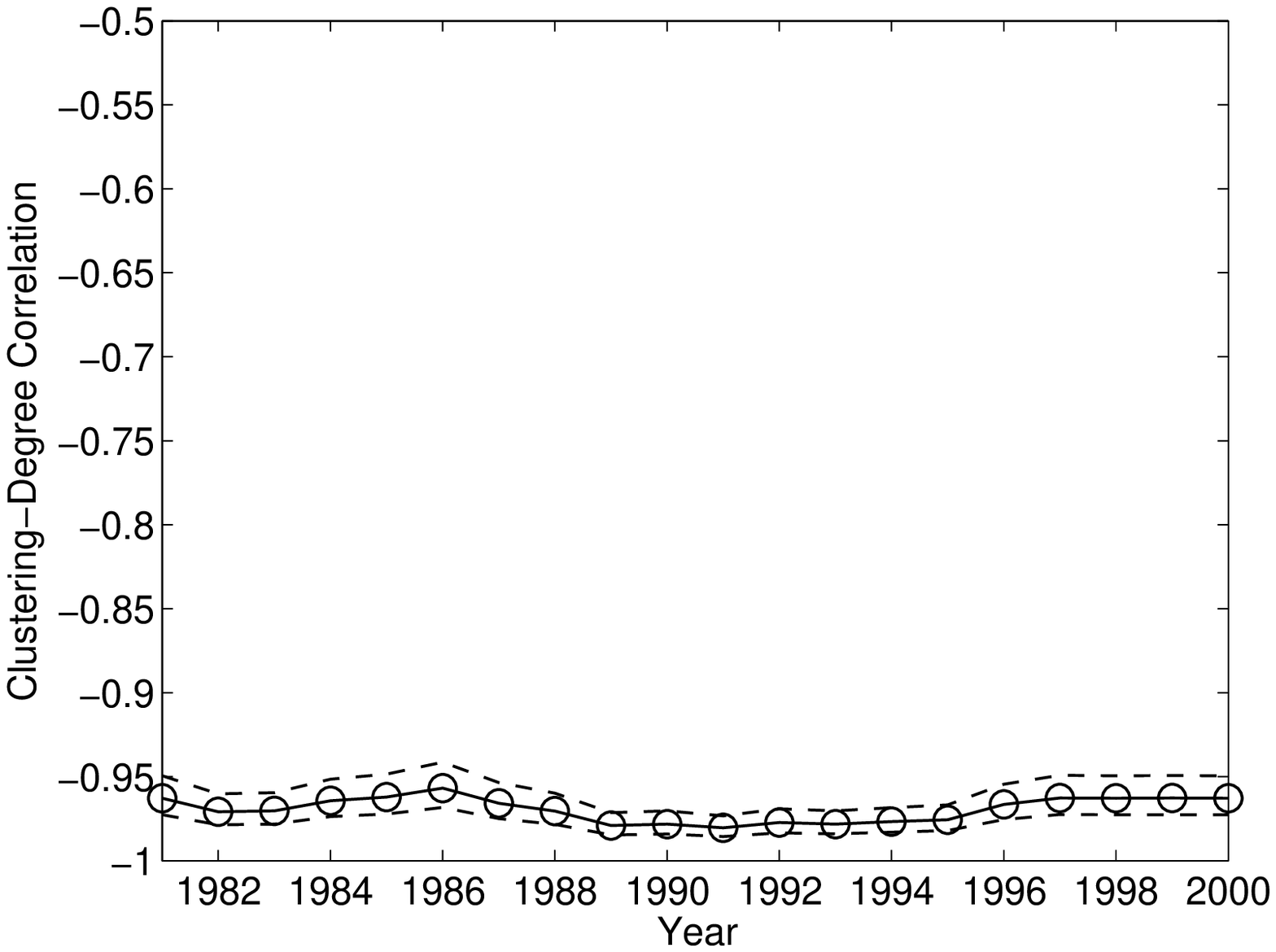}
    \end{minipage}
    \begin{minipage}[t]{6cm}
    \includegraphics[width=6cm,height=4.5cm]{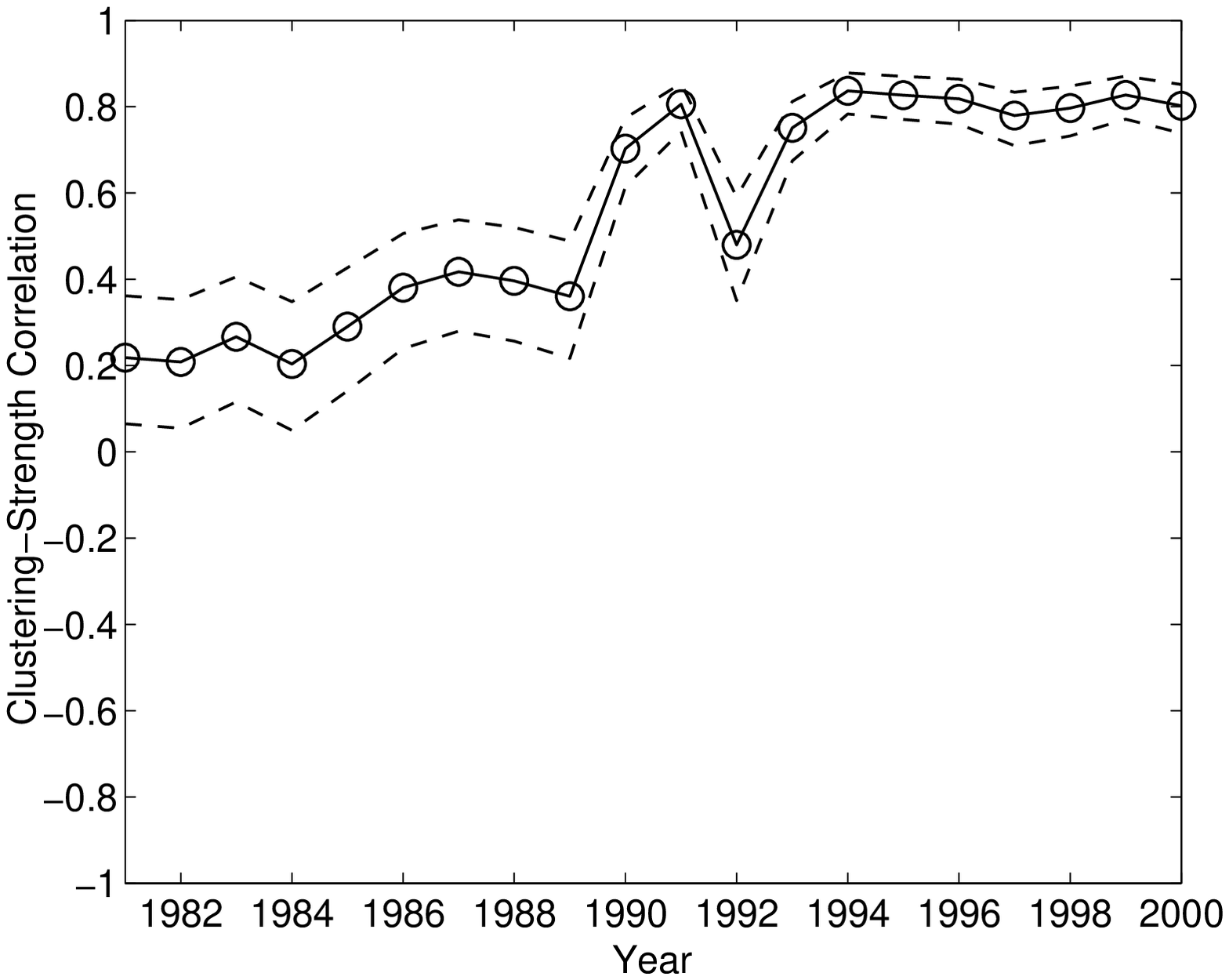}
    \end{minipage}
    \begin{minipage}[t]{6cm}
    \includegraphics[width=6cm]{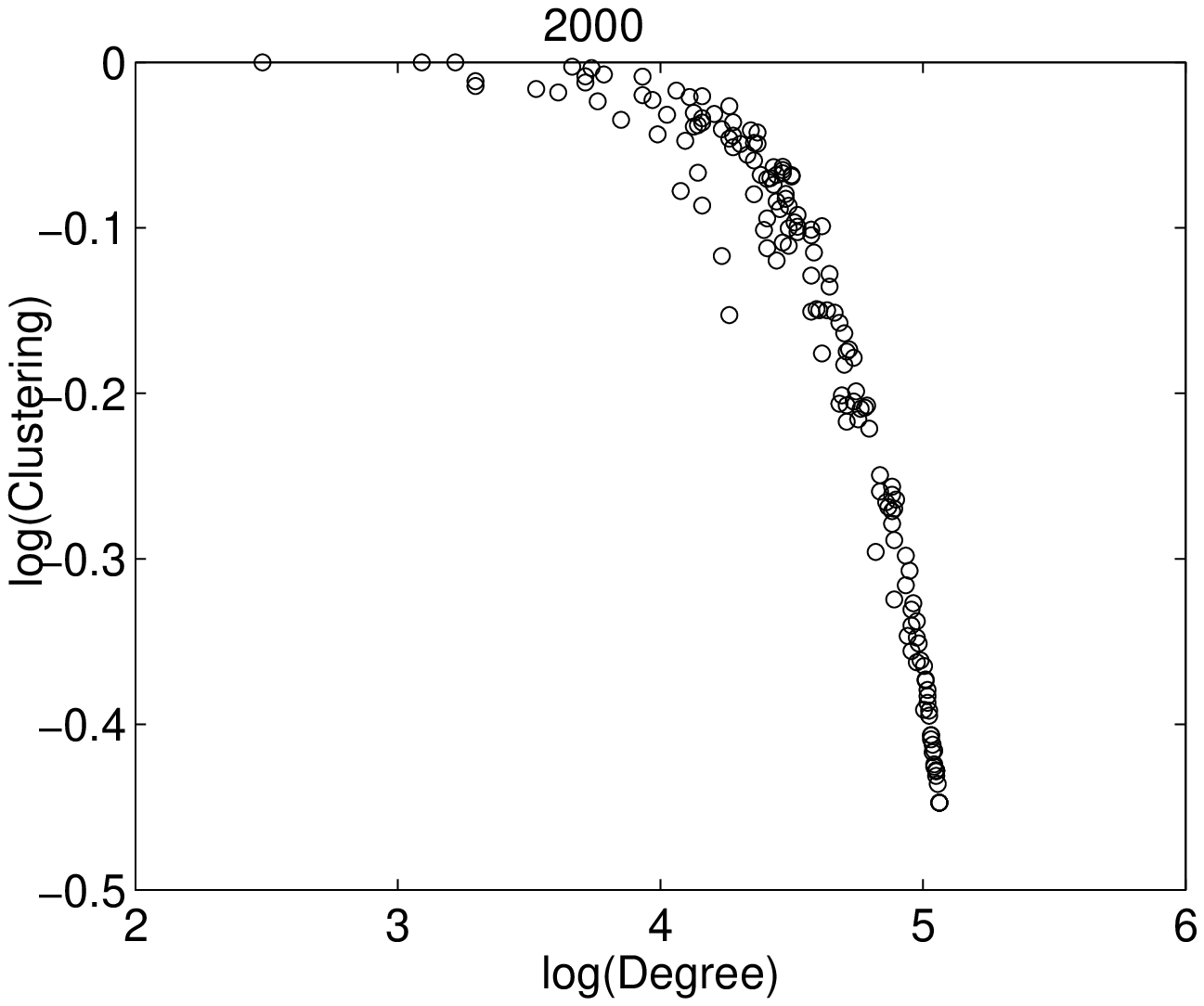}
    \end{minipage}
    \begin{minipage}[t]{6cm}
    \includegraphics[width=6cm]{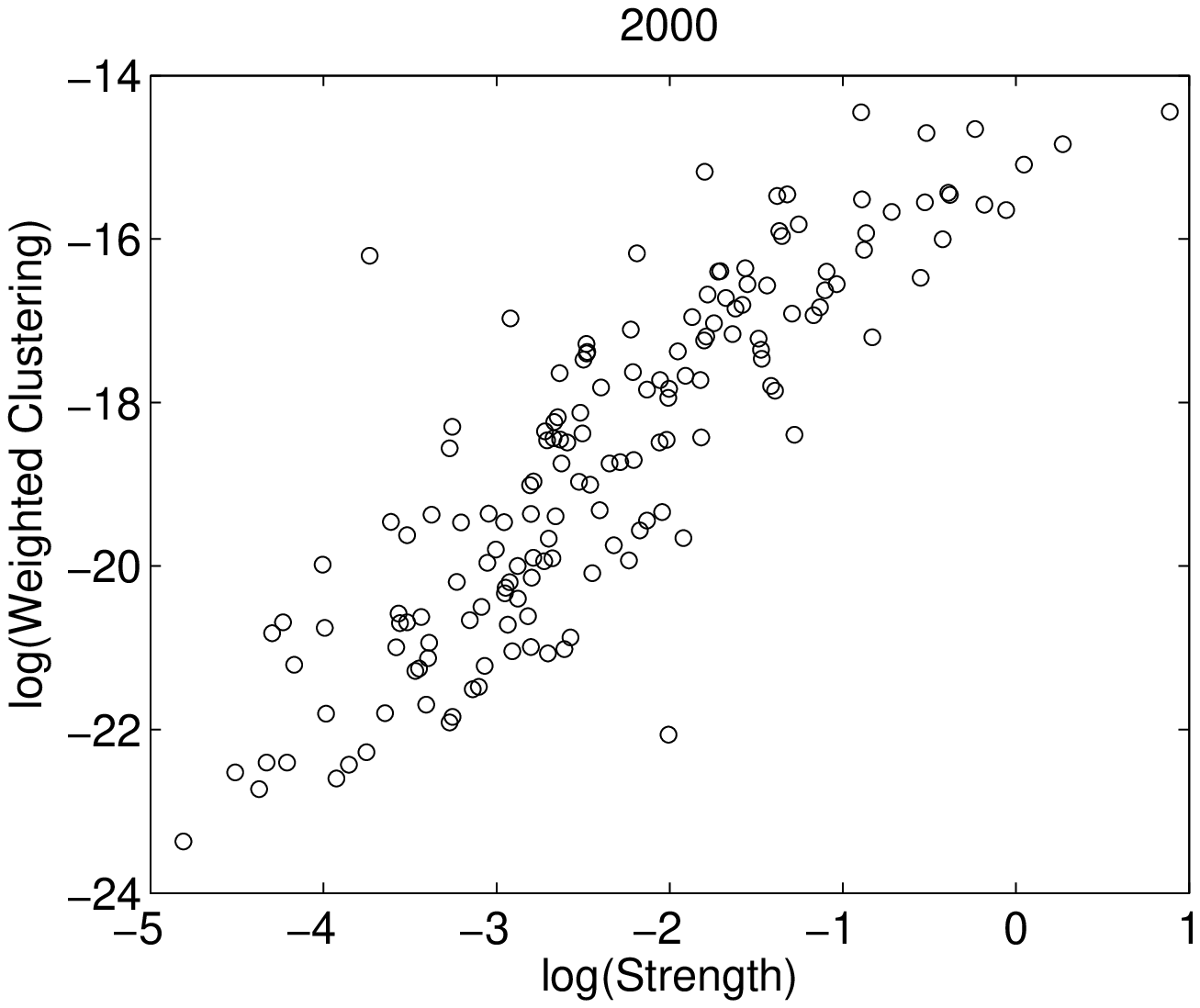}
    \end{minipage}
    \caption{Clustering analysis. Top-Left: Correlation between BCC and node degree over
    time. Top-Right: Correlation between WCC and node strength over time.
    Bottom-Left: BCC-ND (logs) scatter plot in year 2000. Bottom-Right: WCC-NS (logs) scatter
    plot in year 2000. Dashed lines in top panels: 5\% and 95\% confidence intervals.}
    \label{Fig:Clustering}
    \end{figure}

Finally, we address the issue whether SF2 is confirmed by a WNA.
According to Refs. \cite{Garla2005,SeBo03}, countries that in a
binary WTW hold more trade partners are typically associated to
lower clustering coefficients. This means that any two partners
$(h,k)$ of a given country $i$ are not very likely to establish a
trade relationship (i.e., $a_{hk}$ is likely to be zero). Again, a
BNA confirms these results for our data. From a binary perspective,
there exists a -0.96 correlation (stable over time) between BCC and
ND (Figure \ref{Fig:Clustering}, top-left panel). In fact, the
average BCC is very high (about 0.8 on a $[0,1]$ scale). A scatter
plot of BCC vs. ND in year 2000 further corroborates SF2 (Figure
\ref{Fig:Clustering}, bottom-left panel).

Nevertheless, a WNA gives here the opposite result. If viewed as a
weighted network, the WTW displays an increasing, positive, and
significant correlation between WCC and NS (Figure
\ref{Fig:Clustering}, top-right panel), which results in
upward-sloping WCC-NS scatter plots (Figure \ref{Fig:Clustering},
bottom-right panel). Average WCC is actually very low (about 1E-03
on the same $[0,1]$ scale). Therefore, once the existing
heterogeneity of trade relationships is taken into account through a
weighted approach, one finds that countries holding more intense
relationships are \textit{more likely} to form strongly connected
trade triangles. In other words, trade clubs (or cliques) are
typically the case in the WTW.

\section{Concluding Remarks} \label{Section:Conclusions}
In this paper we have performed a WNA to investigate the topological
properties of the WTW in the period 1981-2000. We have studied
network connectivity, assortativity and clustering, and we have
compared our weighted-network results to those obtained using a BNA.
We have shown that the picture stemming from a WNA is substantially
different from that obtained using a BNA. Our results are summarized
in the following table:

  \begin{center}
    \begin{tabular}{c|c|c|c}
    \hline \hline
      & Connectivity & Assortativity & Clustering \\
    \hline
       & Highly Connected &  & Highly-Clustered \\
      BNA & $+$ & Strongly & $+$ \\
       & Bimodal ND  & Disassortative  & Negative CC-ND \\
       & distributions &  & Correlation \\
    \hline
       & Weakly Connected &  & Weakly-Clustered \\
      WNA & $+$ & Weakly & $+$ \\
       & Skewed ND & Disassortative  & Positive CC-ND\\
       & distributions &  & Correlation \\
    \hline \hline
    \end{tabular}
  \end{center}

The above findings support the idea that accounting for the
heterogeneity of interaction intensity in networks is crucial to
better understand their complex architecture \cite{Barr05,Bart05}.
In our application to the WTW, for example, the binary
representation of the WTW leads to a highly-connected graph, where
all links have the same impact on the resulting statistics. Almost
by definition, this implies very large values of ANND and clustering
coefficients for the majority of nodes. Therefore, the computation
of correlation patterns is somewhat biased by this high-connectivity
level. However, different links carry a very different interaction
intensity in the WTW: the majority of links are indeed associated to
low import/export flows (e.g., as a percentage of exporter GDP).
This is true irrespective of the particular method one may use to
weight the links. By exploiting this additional information, a WNA
allows one to better grasp the underlying topological structure of
the network under study.

\end{document}